\documentclass[a4paper,10pt]{article}
\usepackage{graphicx}
\usepackage{amssymb}
\usepackage{float}
\usepackage{graphicx}

\usepackage{graphicx}
\usepackage{amssymb}
\usepackage[utf8]{inputenc}
\usepackage[english]{babel}
\usepackage{xcolor}

\usepackage[utf8]{inputenc}
\usepackage[english]{babel}
\usepackage{xcolor}

\begin{document}
2021/oct/29
\begin{center}
  {\bf \huge Novel String Field Theory and Bound State, Projective
    Line, and sharply 3-transitive group}\\
 \vspace{2mm}
        {\large {\bf H.B. Nielsen}, Niels Bohr Institut\footnote{Speaker:
          Talk presented in 24th Bled-Workshop ``What comes
          beyond the Standard Models''Bled, July 3-11, 2021, Slovenia} \\
  E-mail: hbech@nbi.ku.dk\\
{\bf Masao Ninomiya}, Yukawa Institute for Theoretical Physics,\\ Kyoto University, Kyoto 606-0105, 
Japan\\ and\\
Yuji Sugawara Lab., Science and Engineering,\\
 Department of Physics Sciences, Ritumeikan university\\
E-mail: msninomiya@gmail.com
}
\date{``Bled'' , July , 2021}
\end{center}

\begin{abstract}
  Using ideas from our long studied Novel String Field Theory we consider in
  this article a bound state of infinitely many constituents, something that
  at least very approximately could mean a hadron, since hadrons have typically
  very many constituents. Our main point, so far, is to speculate that there
  should be a very high degree of symmetry between the many consituents,
  since the constituents 
  behave similarly at different places in the bound state. We assume
  speculatively that there is a group represented sharply 3-transitively
  as permutation of the constituents; one can namely only have {\em finite}
  number of elements sharply n-transitively permuted for n larger than 3.
  The scattering of such bound states will in the zero Bjorken $x$ limit
  (which is suggeted) only occur by exchange of parts of the system of
  constituents, quite like in our Novel String field Theory
  the ``objects'' are exchanged in bunches. The cyclically
  ordered chain of objects in this Novel String Field Theory are identified as
  a projective line structure. Also a p-adic field is a natural possibility.
  \end{abstract}



\newpage

\section{Introduction}
{\bf \huge }
\subsection{ Bound State with Infinitely Many Constituents  }

For a bound state\cite{BS} of infinitely many constituents you would at first
expect that the momentum of such a bound state would be shared evenly, so
that each constituent would have a negligible part of the total momentum of the
bound state. This is of course not safe, since a small part of the constituents
might carry the bulk of the momentum, but then the majority of constituents
would carry even less. One usually talks about a Bjorken-$x$\cite{Bjorken6}
defined for each
constituent and denoting the (average) fraction of the bound state momentum
carried by that constituent.


  {\bf Scattering of Constituent on Constiuent Not Important for Many
    Constituents}

  If the single constituents carry only infinitesially small fraction of the
  momentum of the bound state, the scattering of one constituent in one
  bound state with one in another bound state would not be much connected
  to the scattering of the two bound states.

  Rather {\bf scattering of bound states on each other would be dominated by
    one bound state exchnging a bunch of consituents with the other bound
    state.}

  {\bf Scattering by Excange of Constituents}
  \begin{figure}
  \includegraphics{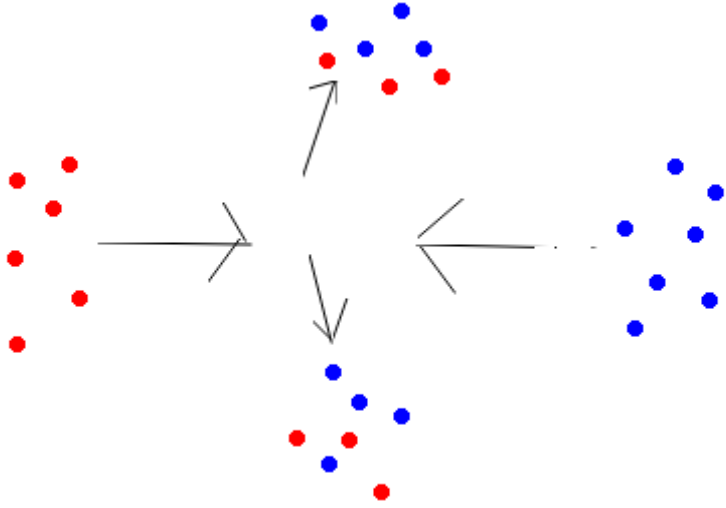}
  
   \caption{{\bf Scattering by Excange of Constutents}
    This figure illustrates the scattering of two bound states with
    (inifinitely) many constituents marked in the one as blue and in the
    other bound state as red. After the scattering there appears again two
    bound states, but now notice that both of them have partly blue-marked
    and red-marked constituents. This we may call an ``exchange of parts of
    the constituents''-scattering.}
   \end{figure}

  {\bf A Motivation for People Interested in making Higher Dimensional
    Theories:}
  It is wellknown: You do not have genuine renormalizable quantum field
  theories in
  higher than 3+1 dimensions.

  However Exception: A very bad scalar theory with $\phi^3$-interaction in up to
  5+1, and {\bf completely free theories}.
  
  \subsection{Novel String Field Theory of Ours}
  
According to the Novel String Field Theory of ours\cite{self2, self8}, which
we shall
  go a bit into later, (super){\bf string theory can be considered a
completely free theory}!  
  
      {\bf ``exchange of parts of constituents''-scattering known from our
        Novel String Field Theory}

      In our novel string field theory, on which we worked much earlier, the
      strings are - somewhat similarly, but differently, to/from C. Thorn's
      \cite{Thorn}
      string bit resolution of the string into ``bits''
      (consituents) - described by means of ``objects''. After scattering of a
      couple of strings (almost, except for a null-set) all the objects
      from the initial strings are either refound in the final state strings
      or recognized as having been annihilated. Although we can consider the
      ``objects'' constituents, they do not scatter on each other, but rather
      do not interact at all.

    \section{Motivation}
   
      \subsection{ Motivation and Plan}
      \begin{itemize}
      \item Interpret the great feature of string theory to be that it is
        indeed  - in our Novel String field theory - a basically free and
        therefore solvable theory, so that even no divergence problems appear.
      \item Ask if we can generalize such a string theory, still clinging
        the idea that the ``constituents'' (identified with our ``objects'')
        do not interact under the scattering of the strings (identified
        as ``bound states'').
      \item Thereby getting e.g. meaningfull (~renormalizable) theories in
        higher than 3+1 dimesnions.
        
        \end{itemize}
   
    \begin{frame}
      
      \subsection{This work especially: Generalize M{\"o}bius Transformations}

      Our series of objects making up so to say an open string in our
      novel string field theory are organized in what we call a cyclically
      ordered chain, which is topologically a circle. It has in fact a
      ``natural'' symmetry under a M{\"o}bius group, as we shall explain, and
      can also be considered a projective line (meaning a line as in
      projective geometry, in which one adds to the lines an extra ``point at
      inifinity'', so that the line topolgically becomes a circle rather
      than a usual line).

      As a major part of the presentation we like to seek to go back from a very
      general group being analogous to the M{\"o}bius group to see to what
      extend we can reconstruct projective line for some field (in the sense
      of the algabraic structure with unit element and invertibility for both
      a multiplication and an addition).
    \end{frame}
    \begin{frame}
      {\bf Reminder of: M{\"o}bius transformations in Veneziano model and
        string theory\cite{Kaku3}}

      From very early times in string theory and Veneziano model theory
      the M{\"o}bius transformations has shown up. In fact physicists\cite{KNV} were so
      kind as to call the variable in the formulation of the Veneziano model
      with some extra variables so that the formulation became precisely
      invariant under M{\"o}bius transformations Koba-Nielsen variables.
      
    \end{frame}
    \begin{frame}
      
      {\bf What is M{\"o}bius transformations ?}
      A priori the M{\"o}bius transformations are defined as transformations
      of the extended complex number set ${\bf C}\cup\{\infty \}$ of the
      complex numbers with a number $\infty$ added, a set equivalent to the
      complex projective line ${\bf CP}^1$ given by the transformation
      function
      \begin{eqnarray}
        z\rightarrow f(z) &=& \frac{az+b}{cz+d},
      \end{eqnarray}
      but shall in the present article be more interested just in the
      real number version transforming only ${\bf R}\cup \{ \infty \}$,
      and with the constants $a,b,c,d$ being real numbers.
    \end{frame}
    \section{Novel SFT}
    \begin{frame}
      
      {\bf How we thought in our Novel String Field Theory in the present
        Articles:}
      We used the splitting of the position variable field on the
      string into left and right-moving parts
      \begin{eqnarray}
        X^{\mu}(\sigma, \tau)&=& X_R^{\mu}(\tau-\sigma)+ X_L^{\mu}(\tau+\sigma),
      \end{eqnarray}
      where $\sigma$ is the ``spatial'' coordinate enumerating the points
      along the string and $\tau$ a ``time'' for the single string, both
      arranged in the conformal gauge, meaning they have been partially gauge
      chosen so that the Lagrangian simplified to a usual 1+1 dimensional
      massless scalar for each value of the external index $\mu$ enumerating
      the imbedding space dimensions 25+1. 
    \end{frame}
    \begin{frame}
      
  \subsection{ Crucial Feature of Our Novel String Field Theory, Use $X_R$ and
        $X_L$.}
      
      Our approach was to discretize into small pieces - analogous to the
      string bits by Charles Thorn, who used the full $X$ - in the varibles
      on which these $X_R$ and $X_L$ only depends, namely
      $\tau-\sigma$ and $\tau +\sigma$ respectively. This means that {\bf we}
      contrary to C. Thorn discretize into pieces variables which are not a
      priori physically enumerating the material of which the string consists,
      but a priori could be just formal parameters enumerating some degrees of
      freedom of the system(=the string). Therefore a priori we could not be
      sure if the ``objects'' corresponding to the small pieces in variables
      $\tau -\sigma$ or $\tau+\sigma$ can be considered ``constituents''.
      
    \end{frame}
    \begin{frame}
      
      {\bf By Changing Physical Interpretation a bit the ``Objects'' may be
        Constituents},though.

      A priori the ``objects'' are  associated only with half the degrees of freedom
      of a string bit - namely only the right or the left moving d.o.f. - are
      {\bf not} genuine constituents. If you speculate that {\bf the string
        is just a smart way of looking at it, but not neccessarily the only
        way,} then we may speculate physically to split up a string bit
      (as by C. Thorn) into two physically seperate objects, a right and
      a left. Since the two, when interpreted as the ``objects'' do not
      interact,
      are not really needed to be considered the same constituent, we can then
      make the physical speculation, or interpretation rather, that the two
      ``objects'' for same string bit are two quite independent
      {\bf constituents}.

      So we are allowed to {\bf take it that the ``objects'' are
        constituents.}
      
    \end{frame}
    \subsection{Objects describing strings} 
    \begin{frame}

      \begin{figure}
      \includegraphics[scale=0.8]{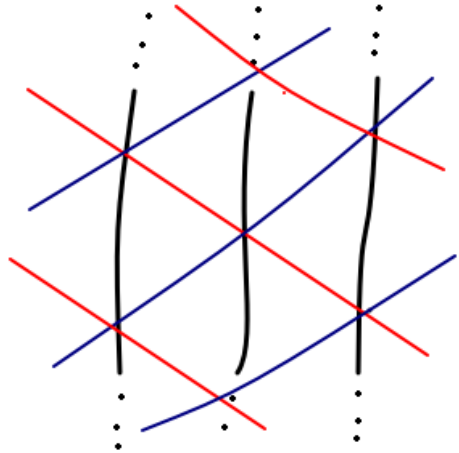}
     \caption{ How strings are seen as crossing places of the objects 
       being Constituents Representing Strings Locally.}
    \end{figure}
    
      The figure shall illustrate how two kinds of ``objects'' denoted
      by red and blue colored lines telling their path through space
      (respectively R and L), when flowing through space in long series
      - infinitesimally close to the neighbors -can represent/look like a
      string moving with lower velocity.
\vspace{1 mm}

The ``objects'' move with velocity of light - actually they are free so they
never change even direction -, but the string seemingly there move typically
slower. The string at one moment is just where the objects meet at that moment.
\vspace{1mm}

The strings are just some way of seeing the objects.
    \end{frame}
    
    \begin{frame}
      
      \subsection{Philosophy of Looking at String Theory in this Talk:}

      {\bf String Theory is a successful theory in higher dimensions
        because it is actually} - according to our Novel String Field Theory -
      {\bf a free theory}, so that it is, one can say, renomalizable even in
      higher dimensions.
     \vspace{1mm}
     The `` objects'' are
     free massless particles.
    \end{frame}
    
    \begin{frame}
      
    {\bf String Versus Novel Object Chains?, Is there a truth?}

    We claim that there seemingly are {\bf two different ways} of imagining the
    strings in string theory:

    \begin{itemize}
    \item{1,} The {\bf strings} are the true physical objects.
      
      \item{2.} The {\bf chains of ``objects''} are the true physical objects.
    \end{itemize}
    
    You may of course claim, that if we are right that the two ways of
    looking at it  are equivalent, then both are right!

    But you could also begin to find argument, that one viewpoint is better or
    more true than the other one:

    In a moment we shall give a couple of weak
    arguments, that the {\bf chains of objects are more true!}

\end{frame}
      
      {\bf Mysterious in String Theory: Cross sections for End and String
        crossing are same order of magnitude?}
      \begin{figure}
        \includegraphics{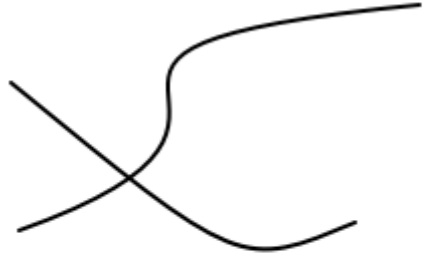}
        \caption{On this figure you should imagine that the two strings
          truly cross each other in the sense of going through the same point.
          This type of hitting each other of two strings is more likely
          than that they should just hit each other just with the end points} 
        \end{figure}

    \begin{frame}
      
      {\bf You expect End hitting much more unlikely than hitting of
        proper string bulk}
      
      \begin{itemize}
        \item
      For two sticks or strings you expect the cross section for that they hit
      to be of the order of the product of their length.
\item
      But two genuine point particles will have in principle zero
      crossection for hitting each other.
      \end{itemize}

      Thus we are forced to make a 
      {\bf Conclusion:}

      Something wrong with string interpretation!

      \end{frame}


      {\bf Can Vacuum Extensions of String Tails Solve Mystery by one string
        having an end common with another without knowing}
      \begin{figure}
      \includegraphics
      [scale=0.8]
      {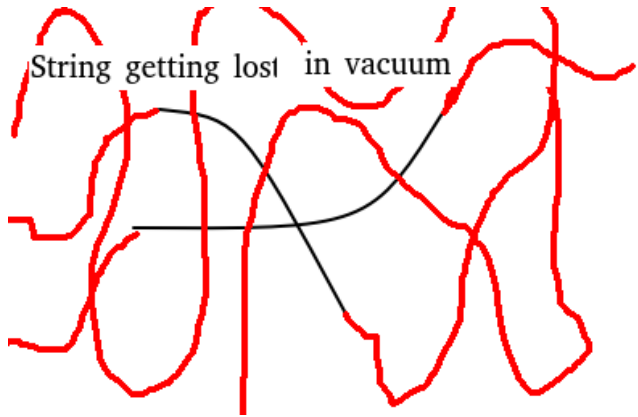}
      \caption{This figure illustrates how one may think, that all the strings
        continue into being virtually present in the vacuum, so that although
        we at first thinks of strings being like particles located to a place
        and moving around, then in fact these phenomenological strings have
        indeed tails continuing out as virtual present string-material
        present virtually in the vacuum (which is of course in all quantum
        (field) theories a very complicated state). We call this possible
        phenomenon of virtual string pieces in vacuum ``Strings lost in
      vacuum''.}
      \end{figure}

    \begin{frame}
      
      {\bf Scattering of two circlar chains (of ``objects'') allways goes with
        two local interactions of the chains}
\begin{figure}      
  \includegraphics[scale=0.8]{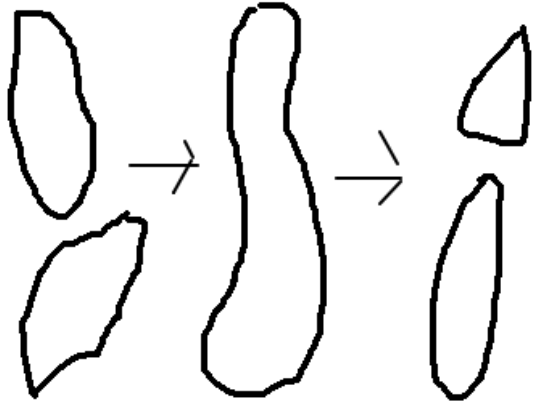}
  \caption{Here we illustrate how two cyclically ordered chains of objects
    (to the left) becomes two other cyclically ordered objects (to the right)
    via two steps of local modification. First - illustrated by the
    left-most of the two arrows - the two cyclically ordered chains
    on the left side by a single local combination becomes the single
    cyclically ordered chain illustrated in the midle. Next this single
    cyclically ordered chain touches itself and thereby split into the two
    cyclically ordered chains illustrated on the right-most.} 
\end{figure}
    \end{frame}
    
    \section{Generalizing}

    \begin{frame}
      
      {\bf It Would be Wonderful to Generalize String Theory, Now we say
        it is Free}

      Basically as soon as you calculate scatterings by approximating that
      all constituents continue without interacting, you are in our
      present sense generalizing string theory.

      So bound states of very many constituents so as each of them
      having very little momentum share are scattering as a generalization of
      the strings seen as composed from objects.
    \end{frame}

    \section{Symmetry}

    \begin{frame}
      
      {\bf Symmetry}
      
      If a bound state or an almost bound state consists of infinitely
      many constituents, then one will, unless there are infinitely many
      types of particles, expect that most of the constituents are in many ways
      very similar in their way of sitting in the bound state.

      {\bf One thus expects a large amount of symmetry between the
        constituents.}

      An idea to implement this expectation is to postulate a group of
      transformmations of the constituents into each other, under which
      the ``structure'' of the bound state is invariant.
    \end{frame}
    
    \begin{frame}
      
      {\bf Much Transfromations / Much Symmetry High n Transitivity}

\begin{figure}      
  \includegraphics{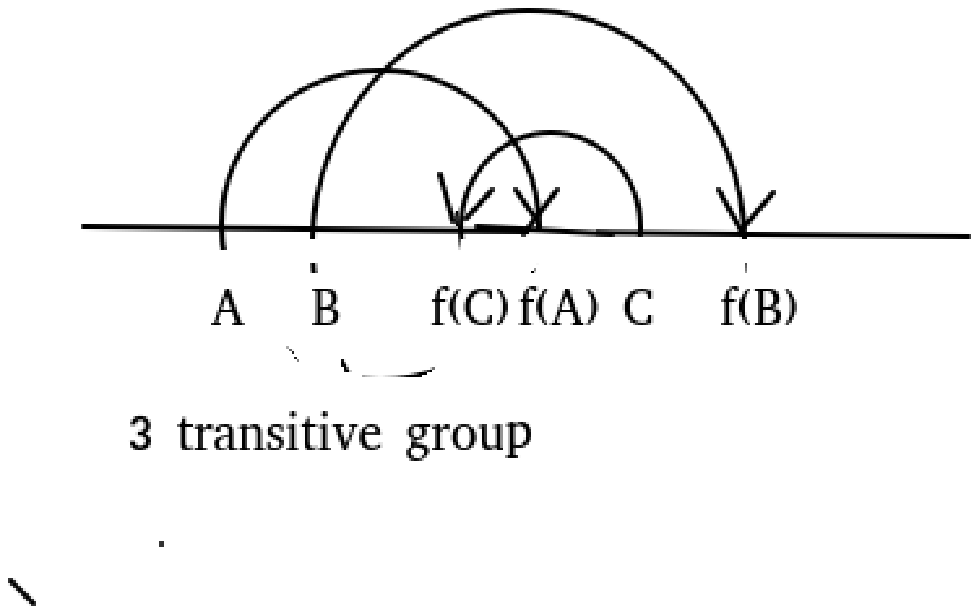}
  \caption{The horizontal line here illustrates the set on which a group $G$
    acts shaply 3-transitively meaning that there is a
    unique (that is what ``sharply'' means) group element $f$ mapping the
    three points $A$, $B$, and $C$ into three given points
    $f(A)$, $f(B)$, and $f(C)$ (that is what means it is 3-transitive).} 
      \end{figure}
    \end{frame}
    
    \begin{frame}
      
      \subsection{3-transitive group action}

      We say the group $G$ acts on the space $S$ $n$-transitively, when
      you for every $n$ points $A$ ,...,$ K$ can find a group element $f\in G$
      transforming these $n$ into $n$ prescribed image points\cite{M4,Hall}. We
      call it ``sharply'', when the group element achieving that is unique.

      \begin{eqnarray}
        \hbox{Any } f&\in& G\\
        \hbox{acts } f&:& S \rightarrow S\\
        \hbox{ For any $n$ points $A, ...,K$}&&\\
        \hbox{and another set of $n$ points } A', ...,K'&&
        \hbox{ there exists }f\\
        \hbox{so that } f(A)&=&A'\\
        &\vdots &\\
        f(K)&=& K'
        \end{eqnarray}
    \end{frame}

    \begin{frame}
      
\subsection{Zassenhaus theorem, that is  a not quite true theorem on
        3-transitive transformations}
      
      A mathematical article by Katrin Tent,
Advances in Mathematics
Volume 286, 2 January 2016, Pages 722-728,
begins:

      `` The finite sharply 2- and 3-transitive groups were classified by
Zassenhaus in\cite{Z14}
in the 1930's and were shown to arise from
      so-called near-fields. They essentially look like the groups of
      affine linear transformations $x\rightarrow ax+b$ or Moebius
      transformations $x\rightarrow \frac{ax+b}{cx+d}$  , respectively.''
    \end{frame}
    
    \begin{frame}
      
      \subsection{Our own Zassenhaus-like
        theorem:}

      Thinking instead Zassenhaus finite groups on infinite ones:
      
      A set on which transforms a group in a sharply 3-transitive way
      will be a projective line corresponding to some field $F$ and the
      transformations under the group will be M{\"o}bius transformations
      $x\rightarrow \frac{ax+b}{cx+d}$ with the variable enumerating the
      points on the ``projrctive line'' $x$ as well as the constants
      $a, b,c,d$ of the transformation(group element) belong to the field
      $F$.
    \end{frame}
    
    \begin{frame}
      
      {\bf Planning to ```derive'' this our own  theorem, like Zassenhaus}

      We should at least reconstruct the field of real numbers ${\bf R}$
      in the case we consider the M{\"o}bius transformations of the
      real projective line ${\bf R}\cup {\infty}$ as the sharply 3-transitive
      group of transformations.
    \end{frame}
    
    \begin{frame}
      
      {\bf First step in Reconstructing the Field $F$ from the Group of
        sharply 3-transitive transformations}
      
      Choose a point in the set $S$ being transformed sharply 3-transitive
      under the group $G$ and call it $\infty$. Then look for the subgroup
      $G_1$ of the group $G$ consisting of the elements in $G$ with only
      one fixed point in $S$, being $\infty$, (and of course also the unit
      element in $G$)

      {\bf The idea is to identify the subgroup $G_1$ having $\infty$
        as the only fix point with the additive group of the field $F$ to
        be found.} The group multiplicaion in $G_1$ inherited from $G$
      of course shall be written with $+$.

      (Say $y, z\in (G_1, *)$, then $y*x=y+z$).

      (Here $*$ is the group multiplication in $G$.)
    \end{frame}
    
    \begin{frame}
      
      {\bf Second Step in Derivation, Identify Scalngs from Two Fix-point
        Transformations}

      Next we notice that by requiring just one more fixed point than the
      $\infty$ we get (at least in the true M{\"o}bius case) a group of
      scalings of the ``numbers'' (the elements in $G_1$ ) around a
      certain number. We might call the second fix-point $0$ and
      a similarity transformation of $G_1$ (the subgroup with
      one fix-point) by one $m$ in the group leaving $0$ and $\infty$ say
      $G_2$,(so $m\in G_2$) say the similarity transformation
      \begin{eqnarray}
        y\in G_1 &\rightarrow& m*y*m^{-1}\in G_1\\
        \hbox{would be called } y&\rightarrow& m\cdot y.
      \end{eqnarray}
      
      This would first be a multiplication with an $m\in G_2$.
    \end{frame}
    
    \begin{frame}
      
      {\bf Third step, Get Identification of $G_2$ with $G_1$ by Selecting
        Point in $S$ to call $1$}

      A priori the subgroup $G_1$ leaving $\infty$ and no other points in $S$
      invariant, is of course different from the subgroup $G_2$ of elements
      in the 3-transitive transformation group of $S$, which we called $G$
      having only two invariant points $\infty$ and $0$ $\in S$.

      We may, however, choose a third point $1\in S$
      (just in a few lines we call it explicitly $1_{in \; S}$, because we want
      to use the notation $1$ also for an element in
      $G_1$, which we must then call $1_{in \; G_1}$ to distinguish)
      different from the
      two points $\infty$ and $0$ $\in S$ and define a correspondence:
      \begin{eqnarray}
        y\in G_1 &\sim& m_y = m \hbox{ so that } y= m\cdot 1= m*1*m^{-1} 
      \end{eqnarray}
      (here we needed a $1\in G_1$, but we can make a corresponding $1$ in
      $S$ as $1_{in \; S} = 1_{in \; G_1}(0)$. Remember that $1_{in \; G_1}$ is indeed
      an element in subgrouop $G_1$ of $G$ and thus a map $1_{in \; G_1} :S
      \rightarrow S$ so that it makes sence to take the image of an
      element in $S$, namely
      $0 \in S$).
      
    \end{frame}

    \section{Why just sharply 3-transitive?, A good question.}

    Above we went to consider just sharply 3-transitively acting symmetries
    as the symmetry required for the system of constituents of the bound state
    with very many consituents and very much symmetry. It was done in the spirit
    that 3 was a high number, when we talk about sharp transitivity. But is it
    a high number, and why just 3?

    The answer is that this 3 is indeed the highest transitivity one can have,
    if one wants the set on which the group acts to be infinite. 

    Jordan\cite{M4} classified, that  finite quadruply transitive group
    in which only the identity fixes four letters must be one of the
    following groups: the symmetric group of four or five letters, the
    alternating group of six letters or the Mathieu group\cite{M14} on eleven
    letters.
    In a work by Marshall Hall\cite{Hall}  we find this work slightly
    extended and
    especially implying that {\bf there is no infinite set on which a group
      acts sharply 4-transitively}.

    Thus in our search for a bound state consisting of infinitely many or at
    least more than 11 constituents (the Mathieu group case; we shall not
    here tell exactly what a Mathieu group is) we simply cannot find any
    sharply 4-transitive group acting on it. We must be satisfied at most with
    a 3-transitive.

    \section{Wavefunction?}
    
    \begin{frame}
      
      {\bf No Interaction: What determines the Wavefunction?}

      If we either assume or approximate away the interaction between the
      constituents, then what can determine the wave function?

      Usually the wave function of a bound state in non-relativistic physics
      is given as an eigenfunction of a Hamiltonian
      \begin{eqnarray}
        H \psi(\hbox{``constituent-positions''})  &=&
        E \psi(\hbox{``constituent-positions''})
      \end{eqnarray}
      
      But relativistically one has to use the Nambu-Bethe-Salpeter
      equation\cite{BS},
      which is analogous to this eigenvalue equation for a bound state, where
      $E$ is the enrgy of the bound state.
      {\bf  But if no interaction the
        Hamiltonian $H$ is just 
        trivial (essentially $0$) and of no help!}
      
    \end{frame}
    \begin{frame}
      
      {\bf Nambu-Bethe-Salpeter-equation}
      
      \begin{center}
      \begin{figure}[h]
        \includegraphics[scale=0.3]{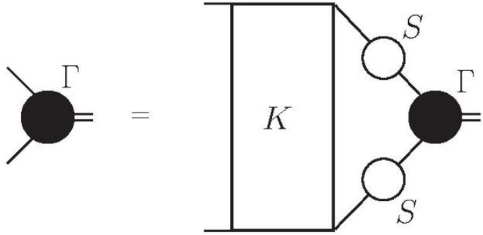}
        \caption{Formal graphic writting of the Bethe-Salpeter equation
          with two constituents.
          The propagators with $S$ are  the propagators for the constituents,
          the black sphere $\Gamma$ is the wave function for the bound state,
          and the block $K$ symbolizes the interaction between the constituents.
          You may consider the Bethe-Salpeter-euation as analogous to the
          eigenvalue equation in the case of non-relativistic constituents.
          It is worth calling attention to, that while we for non-relativistic
          binding are accostumed to only describing the constituents by say
          their momenta (and the energy is only a consequence of that and it is
          positive energy for the constituent,) then here in
          the Nambu-Bethe-Salpeter equation the energy (or equivalently
          it conjugate $x^0$) is also formally a degree of freedom for the
          constituent. And the energy of the constituents can aslo be negative.}
          
      \end{figure}
      \end{center}
      
      
      
 \end{frame}
    \begin{frame}
      
      {\bf Alternatives to make Wavefunction (meaningful)}
      
   \begin{itemize}
   \item Postulate a new law of nature on initial and even final conditions
     specifying the nature of allowed wave functions (to have e.g. the
     smooth chain character giving the string topologically)
   \item Say that you do not take non-interaction 100\% serious but
     allow very short scale or high energy interactions.

     This would be very realistic in higher dimensions than 3+1.

     In fact couplings of a certain energy scale order of magnitude
     are very weak in higher dimensions (at low energy). So one might at low
     energy in higher dimensions only see the interactions as left over
     influence on the wave functions, but thet would be negligible in
     the scatterings
     (at low energy).
     \end{itemize}
 \end{frame}
    \begin{frame}
      
   {\bf Speculation of mainly  Short Distance Interacting Particles in High
     Dimensions}
   
   \begin{itemize}
   \item From dimensional arguments the interaction between particles in an
     effective quantum field theory in high dimension goes to zero for small
     energy scales. So the particles have only short range interactions.
   \item It is very easy that bound states can exist bound by the short range
     forces; but they will typically have interactions only at short distances
     too and high masses unless mass protected.
   \item If some bound states or ``fundamental'' particle, e.g. from being
     chiral fermions, are mass protected they will of course be massless untill
     their symmetry protection somehow gets spoiled.
   \end{itemize}
 \end{frame}
    \begin{frame}
      
    {\bf Speculation of mainly  Short Distance Interacting Particles in High
      Dimensions (continued)}
    
   \begin{itemize}
   \item But if they are of small extension - like fundamental scale -
     they will effectively not interact from a low energy scale point of view.
   \item Only if we have spatially largely extended bound states, can
     there be appreciable interactions at low enrgy scales.
     \end{itemize}
 \end{frame}
    \begin{frame}

   \begin{center}
   \begin{figure}
     \includegraphics{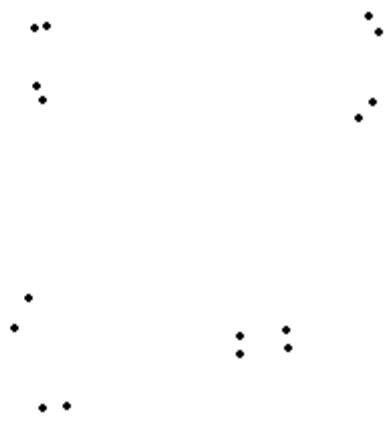}
     \caption{Here we have drawn a presumably very realistic picture of
       a bound state of many particles. Analysing perturbative quantum field
       theory you will typically find that each dressed particle consists of
       at first two consituents - so that the dressed particles are really
       boud states - but since when even further analysed, these will again turn
       out to consist of more constituents each, even the dressed particles
       come to consist of more and more consituents. So a picture with pairs
       forming again pairs and so on is very realistic for such perturbatively
       treated states. Such a picture with pairs forming pairs etc. is also
       very similar to what gets when the field we arrive at happens to be
       the 2-adic field. } 
   \end{figure}
   \end{center}

    \end{frame}

    \section{Hamiltonian ?}

    There is a little not quite acceptable point in our
    ``novel string field theory''
    consisting in, that we stress that there is no development or at least that
    ``objects'' behave freely - in fact the position moves in a trivial
    way if we identify a conjugate variable to what is essentially the momentum
    (called in our papers $J$) - but nevertheless we had also a
    paper\cite{selfhamiltonian}
    in which
    a Hamiltonian appeared, which gave the usual string spectrum. That of
    course cannot both be right: Either we
    have trivial or no development at all or we have the not completely
    trivial Hamiltonian formulated in our variables for the ``objects''.

    Here we must remind of an a bit technical detail in our novel string field
    theory:

    Because variables  the $X_R$ and $X_L$ which we used to take out the
    degrees of freedom for the ``objects'' do not commute with themselves
    taken at neighboring points of the variable such as $\tau-\sigma$ we had
    to after having divided the variable $\tau - \sigma$ into small pieces
    to include in the true representing objects only every second one of the
    pieces. That we formulated by saying we only take as the true objects
    the small pieces with an {\em even } number in the series along in the
    variable $\tau - \sigma$. Then the idea was to instead represent the odd
    numbered pieces as being proportional to the difference of the conjugate
    momenta to the variables on the even places.

    Without going too much in detail with this problem of only using every
    second of the bits into which we cut the variable $\tau -\sigma$ we may tell
    that we obtained a Hamiltonian $H$ which gives the energy of the string
    in say infinite momentum frame expressed by means of the momentum variable
    (which we called $J^{\mu}(I)$ and defined only for even values of the
    piece enumerating integer $I$) and the conjugate position variable
    (which we called $\Pi^{\mu}(I)$ again using only even $I$).

    The crux of this writting down the Hamiltonian, that could give the
    string energy is that not only does it contain the for a free theory
    expected terms going with the square of the momenta, but it had also,
    namely corresponding to the odd pieces in the chain which were of the
    form
    \begin{eqnarray}
      (\Pi^{i}(I+1)-\Pi^{i}(I-1))^2&=& (X_R^i(I+1)-X_R^i(I-1))^2
      \hbox{ (for $I$ odd)}
      \end{eqnarray}
    Such terms look like interaction terms between the neighboring
    even numbered ``objects'' and should not be allowed if we want to
    think of the model for the string as being a construction / a bound state
    made out of {\em non-interacting constiuents}!
    
    Infact the Hamiltonian giving the usual energy of the string takes the
    form, for the Mass square $M^2$:
    \begin{eqnarray}
      M^2 &=& 2P^+P^--\Sigma_{i=1}^{24}(P^i)^2\\
      &=& 2\left(\Sigma_{I=0}^{N-1}J^+(I)\right )
      \left (\Sigma_{I=0}^{N-1}J^-(I)\right )
      \left(\frac{1}{2\pi \alpha'}\right)^2\\
      && - \Sigma_{i=1}^{24}
      \left ( \Sigma_{I=0}^{N-1}J^i(I)\right)^2\frac{1}{(2\pi \alpha')^2}
    \end{eqnarray}
    in infnite momentum frame, where the summation $I$ runs over the
    objects as sitting along the cyclically ordered chain (identified with
    the projective line discretized) and $i$ over the transverse dimenisions,
    while the infinite momentum frame coordinates denoted with index ${}+$
    and the index ${}^-$ take care of the longitudinal momentum and of
    the energy.

    In our works on the novel string field theory we make in order to
    not have too many degrees of freedom in the description with the
    ``objects'' the trick of replacing the odd $I$ objjects present at first
    by an expression in terms of the conjugate momenta for the even $I$
    variables $J^{\mu}(I)$,(for the transverse coordinates $i$ say)
    \begin{eqnarray}
      J^{i}(I)&=& -\pi \alpha' (\Pi^i(I+1)-\Pi^i(I-1)).
      \end{eqnarray}

    In this notation with only the even objects being taken as physical,
    while the odd ones are replaced by the conjugate of the neightbors
    the mass square may be rather written
    \begin{eqnarray}
      M^2&=& \frac{1}{(2pi \alpha')^2}\Sigma_{i=1}^{24}\left (
      N\Sigma_{I=0\; I \hbox{even}}^{N-2}(J^i(I))^2-\Sigma_{K=0 \; K \hbox{even}}^{N-2}
      \Sigma_{I=0\; I \hbox{even}}^{N-2}J^i(I)J^i(K) \right ) \nonumber\\
      && +\frac{N}{4}\Sigma_{i=1}^{24}\Sigma_{I=1 \; I \hbox{odd}}^{N-1}(\Pi^i(I+1)
      - \Pi^i(I-1))^2
      \end{eqnarray}
    For details on this kind of Hamiltonian expessions we refer to our
    work \cite{selfhamiltonian}.
This kind of Hamiltonian expressions 
     obviously has the problem of having the seeming interaction
    between the neighboring even numbered objects.

    But then how can we keep up our claim of free constituents?

    The idea to overcome this seeming contradition in our description of
    our novel string field theory previously has to do with yet a technical
    point to which we must allude:

    We had to impose a condition that the state of the cyclical chain of
    objects to make up the description of an open string should obey
    \begin{eqnarray}
      J^i(I+1)&\approx &- \alpha'(\Pi^i(I+1) -\Pi^i(I-1)) \approx J^i(I-1)
      \hbox{ ( here $i$ odd)}\label{pdx}
      \end{eqnarray}
    where in our a bit stupid notation the momentum of an even object is
    denoted by $J^i(I)$ for $I$ even and the corresponding postion variable
    in our old notation is $\Pi^i(I)$ correspondingly, since the $\Pi$ and
    the $J$ are conjugate. 

    But now the way this approximate relation has to be implemented is by
    the {\em state} of the chain of objects has to be so as to full fill it.

    So this relation relating the vector from one even object to the next
    in position space to the momenta of the two neighboring objects
    is a restriction on the state, one could say an intial state
    condition.

    \subsection{Rewritting the Hamiltonian}

    The crux of the matter of the idea to solve the just above
    explained seeming contradiction is to say:

    {\bf As long as we are only interested in states of the system of
      `` objects'' obeying the equation (\ref{pdx}), we should at
      least approximately be allowed to use this condition (\ref{pdx})
      to substitute parts of the Hamiltonian as using it as an equation,
      and then one can easily compute that we can rewrite the wanted hamitonian
      to totally free one!}

    This then means that indeed we can claim that if we have
    series of infinitely many genuine particles (scalars in the simple case
    of the bosonic string) which are free - they do not interact -
    but are in such a states that they form a chain and further obey
    our constraint (\ref{pdx}), then the free Hamiltonian can for such
    special states be replaced by the usual string Hamiltonian.
    So for the states of relevance we indeed get the usual energy spectrum
    well known for the strings, in spite of the fact that we take the
    cnstituent particles to be genuine particles that do not interact.

    \section{An Idea of a Picture}

    Let us here present the idea, that we should apply the present
    infinite constituent bound state picutre in a world in which the
    genuine physics theory is {\bf higher dimensional} - as e.g. the model
    by Norma Mankoc Borstnik\cite{Norma} - and that we in practice only see
    bound states of very many {\em mass protected particles} such
    as chiral fermions or gauge particles.
    For example in 4 dimensions a chiral fermion is described by spinor
    field $\psi(x)$ imposed
    the restriction
    \begin{eqnarray}
      (\Gamma-1) \psi(x) &=&0\\
      \hbox{where } \Gamma &is& \hbox{the chirality},
      \hbox{ the analogue of $\gamma_5$.}
      \end{eqnarray}
    
    Then at low energies there
    will be effectively no interactions, because the dimensionalities of the
    coupling constants, $\kappa$ say, would be of dimensions mass to
    negative powers. E.g. an intraction could be of the four fermion type
    \begin{eqnarray}
      {\cal L}(x) &
      =& \kappa \bar{\psi}(x)\gamma_{\mu}\psi(x)*
      \bar{\psi}(x)\gamma^{\mu}\psi(x)+ ...\\
        \left  [ \kappa \right ] &=& \left [ GeV^{\hbox{``negative''}} \right ] \\
          \hbox{so } \kappa &\approx& 0 \hbox{ at low energies.}
      \end{eqnarray}
    So bound states of many such mass protected particles would have in the
    effective low energy scattering limit\cite{EFT}  no interactions, just as
    we discussed
    in the present article.

    However, the initial - and even final - conditions for the bound states,
    which in the no interaction approximation need an extra explanation, could
    now be understood as {\bf a left over from the high energy time, when
      the bound states were formed}. That is to say: Once upon a time
    in the time just after Big Bang, say, the coming consituents interacted
    because at high energies, the couplings - like $\kappa$ - with negative
    power of mass dimensions
    would not be quite negligible. This interaction at high energy would
    bring the mass protected particles into bound state structures - once
    cooling takes place - , which
    would then survive into the colder times, when only the low energy
    approximation would be relevant, and the constituents  would effective
    no longer
    interact.

    If it happens that the surviving bound states are of the many
    constituent types, and they would obtain the symmetry properties speculated
    in the present article, they would, if the symmetry becomes the M{\o}bius
    group with {\em real} numbers as the field, become real field
    projective lines, meaning circles topologigally. That is to
    say, we would get the cyclically ordered chains of objects
    described in our earlier Novel String Field Theory\cite{self8}. If one
    would instead
    take the p-adic field, one (presumably) would obtain instead the
    p-adic Veneziano model scheme\cite{padic6}.

    In any case in the here now suggested background model for our infinitely
    many constituent bound states, there would be a usual quantum field theory
    picture in higher dimensions behind, and the model should inherit the good
    physical properties from such a quantum field theory, although it would
    not be a renormalizable one. Only the low energy limit
    with only exchange of consituents interactions would be
    what one might call ``renormalizable''. Really it would in the
    most important case be string theory and that we would rather call
    ``finite'' than renomalizable, but the ``finite'' theories are,
    one could say, included in the renoramlizable ones.

    By taking such an at high energy interacting scheme as the model
    behind our bound states one would get a more solid physical picture and
    could from this picture better understand how to treat e.g. the
    Nambu-Bethe-Salpeter-equation technique. In such a philosophy of the
    high dimensional theory behind one should have a good chance
    using Weinbergs effective field theory\cite{EFT} thinking to see
    that
    the bound state scattering based on the Nambu-Bethe-Salpeter
    formalism\cite{BS}
    would lead to amplitudes acceptable from say axiomatic field
    theory\cite{axiom}
    point of view. It thus looks as an outlook that we are close to
    having a scheme for making models for scattering amplitudes,
    that are physically acceptable.

    \subsection{Scale Symmetry}
    Let us note, that when we in the low energy limit have just the free
    mass protected, say chiral fermions or free ``photons'', then the theory
    is {\bf scale invariant}. There are no dimensionized parameters left.
    So if it is as suggeted string theory the dimensionized parameter
    $\alpha'$ in string theory cannot be in the low energy approximation
    proper, but can only come in via initial and final state conditions.
    That is to say here, that the Regge trajectory slope parameter $\alpha'$
    can only come in by having been inherited from the era, when it was hot
    and cooled down so that the dimensionized parameters such as $\kappa$
    were relevant.

    \subsection{Momentum Fluctuation in ``Unexcited cyclically ordered system
      of objects/constituents''}

    At least for the unexcited bound state, but approximately for all
    in practice occuring bound states in the model, we should have that
    the momentum distribution for the constituents should be invariant
    under the M{\o}bius transformations.

    In the figure
\begin{figure}
  \includegraphics{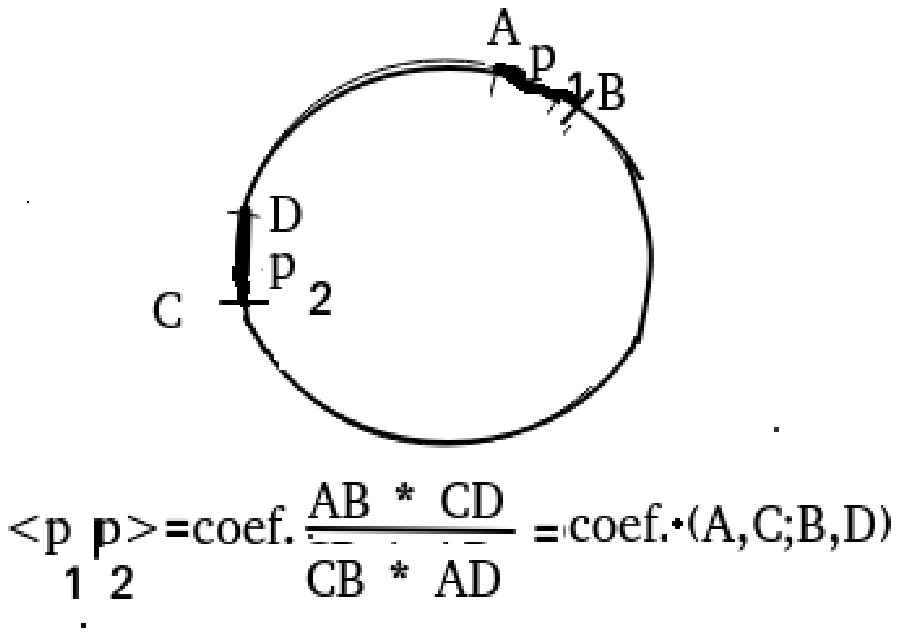}
  \caption{Illustrated are the two small pieces $AC$ and $BD$ with respectively
    momenta $p_2$ and $p_1$ on the projective line drawn as a cicle. We use the
    symbol $(A,C;B,D)$ for the anharmonic ratio, which in terms
    of e.g. distances counted with sign on the circle would be
  $(A,C;B,D)=
       \frac{AB*CD}{CB*AD} = \frac{AB}{CB}:\frac{AD}{CD}$.}
\end{figure}
    we see a drawing of the cyclically ordered chain equivalent to the
    real number projective line. The ``four'' (really we should call
    it d-momentum, because we consider here more than four dimensions in order
    to have the effective disappearance of the couplings) momentum $p_1$ and
    $p_2$ of the collection of the consituents sitting in the projective
    line in the two small (infinitesimally small) pieces $AB$ and $CD$
    respectively have in the ``ground state'' of the bound state the
    correlation
    \begin{eqnarray}
      <p_1p_2> &=& coef. *(A,C;B,D)\\
      \hbox{where } (A,C;B,D)&=&\hbox{``anharmonic raatio''}\\
      &=& \frac{AB*CD}{CB*AD} = \frac{AB}{CB}:\frac{AD}{CD}.\\
      \hbox{or more precisely: }  <p_1^{\mu}p_2^{\nu}> &=&
      g^{\mu\nu}\cdot coef. *(A,C;B,D).
    \end{eqnarray}
    The form given by the anharmonic ratio is specified by the M{\o}bius
    invariance under real numbers, because only the anharmonic ratios
    are invariant for four points. Here the pairs of letters like $AB$ means
    the difference of the field F coordinate for $B$ minus that for $A$.
    However,
    the coefficient $coef.$ is an a priori
    unspecified constant. What the value shall be is only specified by the
    initial and final state information on the bound states we had to impose.
    In the picture of there being a short distance or high energy interaction
    underneath the $coef.$ would inherit from such a short distance interaction.
    In the string theory with the open strings identified with our
    bound states this coefficient would be given by the Regge-slope
    $\alpha'$,
    \begin{eqnarray}
      coef. &\propto& \frac{1}{\alpha'}.
      \end{eqnarray}
    It
    should be in mind that the scale symmetry is only broken by
    these initial and final state informations.

    So the $\alpha'$ energy scale in the Veneziano model in our picture
    comes in via the initial and final state information, only.

    \section{The Cyclic ordering partly violates the
      full M{\o}bius symmetry}

    We have to mention what formally looks like a little problem:

    In our novel string field theory we had to impose the condition
    (\ref{pdx}), which is {\bf not invariant under shift of
      orientation along the cyclically ordered chain}. Formally such
    a condition would break the symmetry under half of the M{\o}bius
    transformations. To have this condition consistent with the symmetry
    we should only keep those M{\o}bius transfromations, which leave
    the orientation along the chain or in the words of the present article
    along the projective line intact. This means that formally the group
    does not act 3-transitively, but only 3-transitively modulo the
    cyclic orientation.

    But from what we could call an estetic point of view this only
    orientation keeping subgroup is quite nice in as far as it indeed lies
    inside the full M{\o}bius group as a topologically seprate part,
    a component.

    It might be needed in our building of the bound states from a high energy
    theory to let this one have sufficient breaking of its symmetries so as
    to deliver such bound states that the orientation gets fixed.
    
    \section{Conclusion}
    
    \begin{frame}
      
      {\bf Conclusion}

      {\bf We have proposed  an approximation applicable hopefully to
        some bound states: that they have so many constituents with so equally
        divided momenta - or better Bjorken $x$'s $\approx 0$ \cite{Bjorken6} -
        that we can ignore
        the scattering of the constituents, when the bound states scatter.}

        (This means the constituents are in the approximation free, and
      thus the bound state not truly bound)

    \end{frame}
    \begin{frame}
      
      {\bf Conclusion Details}
      \begin{itemize}
      \item Requiring High Symmetry in form of 3-transitive symmetry operation
        we expected - like Zassenhaus - the constituents to form a structure
        like a projective line $F\cup \{\infty\}$ for a field $F$.
        The srting is the case $F={\bf R}$ i.e. the field is the real number
        field.(Topologically the projective line is a circle.)
      \item We suggest that such string theory might be used when the
        approximation of many constituents with little momentum each becomes
        good. (of course string theory historically started as attempt to
        describe hadron physics\cite{string})
        \item The p-adic theory\cite{padic6} of Veneziano model is suggestively incorporated.
        \end{itemize}
      
    \end{frame}

\section*{Acknowledgement}

One of us (H.B.N.) acknowledges the 
Niels Bohr Institute for allowing him to work 
as emeritus and for partial economic support.
Also thanks food etc. support from the Corfu conference
and to Norma Mankoc Borstnik for asking for a 
way to get meaningful quantum field theories in higher than 4 
dimensions. The thinking on hadronic like bound states could namely be looked upon 
as an attempt to find such a scheme using bound states as the theory behind the 
particles for which to make the convergent theory.

M. Ninomiya acknowledges Yukawa Institute of Theoretical Physics, 
Kyoto University, and also   the Niels Bohr Institute
and Niels Bohr International Academy for giving him
 very good hospitality
during his stay. M.N. also acknowledges at 
Yuji Sugawara Lab. 
Science and Engeneering, 
Department of physics sciences Ritsumeikan 
University, 
Kusatsu Campus for allowing him 
 as a visiting 
Researcher.


\begin{thebibliography}{99}
\bibitem{Z14}
        Cambridge Togets in Mathematics Book 24 (English Edition);they are
        the five sporadic simple groups called M11, M12, M23, and M24, which
        multiple transitive groups on 11, 12, 22, 23 arzt objects
\end{thebibliography}
\end{document}